\documentclass[10pt,conference]{IEEEtran}

\usepackage{cite}
\usepackage{amsmath,amssymb,amsfonts}
\usepackage{algorithm}
\usepackage{algpseudocode}
\usepackage{graphicx}
\usepackage{textcomp}
\usepackage{xcolor}
\usepackage[hyphens]{url}
\usepackage{fancyhdr}
\usepackage{hyperref}
\usepackage[font={small,it}]{caption}
\usepackage{subcaption}
\usepackage{makecell}

\usepackage{makecell}

\def\frameworkname{Choreographer}
\newcommand{\squeezeup}{\vspace{-0.8em}}
\newcommand{\dontsqueezetoomuch}{\vspace{0.0em}}

\begin{document}

\title{Choreographer: A Full-System Framework for Fine-Grained Tasks in Cache Hierarchies}
\author{
Hoa Nguyen \\ Dept. of Computer Science \\ University of California \\ Davis, CA, USA \\
\texttt{hoanguyen@ucdavis.edu} \\ \\

Jason Lowe-Power \\ Dept. of Computer Science \\ University of California \\ Davis, CA, USA  \\
\texttt{jlowepower@ucdavis.edu} \\ \\

\and 

Pongstorn Maidee \\ AMD \\ Research \& Advanced Development \\ San Jose, CA, USA \\
\texttt{pongstorn.maidee@amd.com} \\ \\
Alireza Kaviani \\ AMD \\ Research \& Advanced Development \\ San Jose, CA, USA \\
\texttt{alireza.kaviani@amd.com} \\ \\
}

\newcommand{\dataneeded}{\textsuperscript{\color{red}{data-needed}}}

\maketitle
\begin{abstract}
In this paper, we introduce \frameworkname{}, a simulation framework that enables a holistic system-level evaluation of fine-grained accelerators designed for latency-sensitive tasks.
Unlike existing frameworks, \frameworkname{} captures all hardware and software overheads in core-accelerator and cache-accelerator interactions, integrating a detailed gem5-based hardware stack featuring an AMBA coherent hub interface (CHI) mesh network and a complete Linux-based software stack.
To facilitate rapid prototyping, it offers a C++ application programming interface and modular configuration options.
Our detailed cache model provides accurate insights into performance variations caused by cache configurations, which are not captured by other frameworks.
The framework is demonstrated through two case studies: a data-aware prefetcher for graph analytics workloads, and a quicksort accelerator.
Our evaluation shows that the prefetcher achieves speedups between 1.08x and 1.88x by reducing memory access latency, while the quicksort accelerator delivers more than 2x speedup with minimal address translation overhead.
These findings underscore the ability of \frameworkname{} to model complex hardware-software interactions and optimize performance in small task offloading scenarios.

\end{abstract}
\section{Introduction}
The new age of domain-specific computing has ushered in a paradigm shift, where custom components are increasingly deployed to optimize performance, power, and energy efficiency for targeted applications.
In modern high-performance computing systems, such components have become indispensable \cite{cohen2024next,mosur2024built,williams2024qualcomm}, addressing challenges posed by the end of Dennard scaling and the limitations of power delivery\cite{radhakrishnan2021power,eeckhout2024focal}. 
Near-data computing (NDC), including processing in memory, effectively reduces energy consumption at the system level~\cite{2018_DM_Bottlenecks}.
However, to achieve optimal efficiency, CPUs typically consist of accelerators and engines tailored for specific domains.
Integrating custom engines directly into CPU dies enables fine-grained interactions between CPUs and specialized engines, which is particularly advantageous for low to medium sized tasks--referred to as fine grained tasks--such as small offloaded and data-triggered operations~\cite{2024Leviathan}.
These tasks often require frequent CPU interaction and are not ideally suited for NDC approaches.
\begin{figure}[t]
\centering
\includegraphics[scale=0.19]{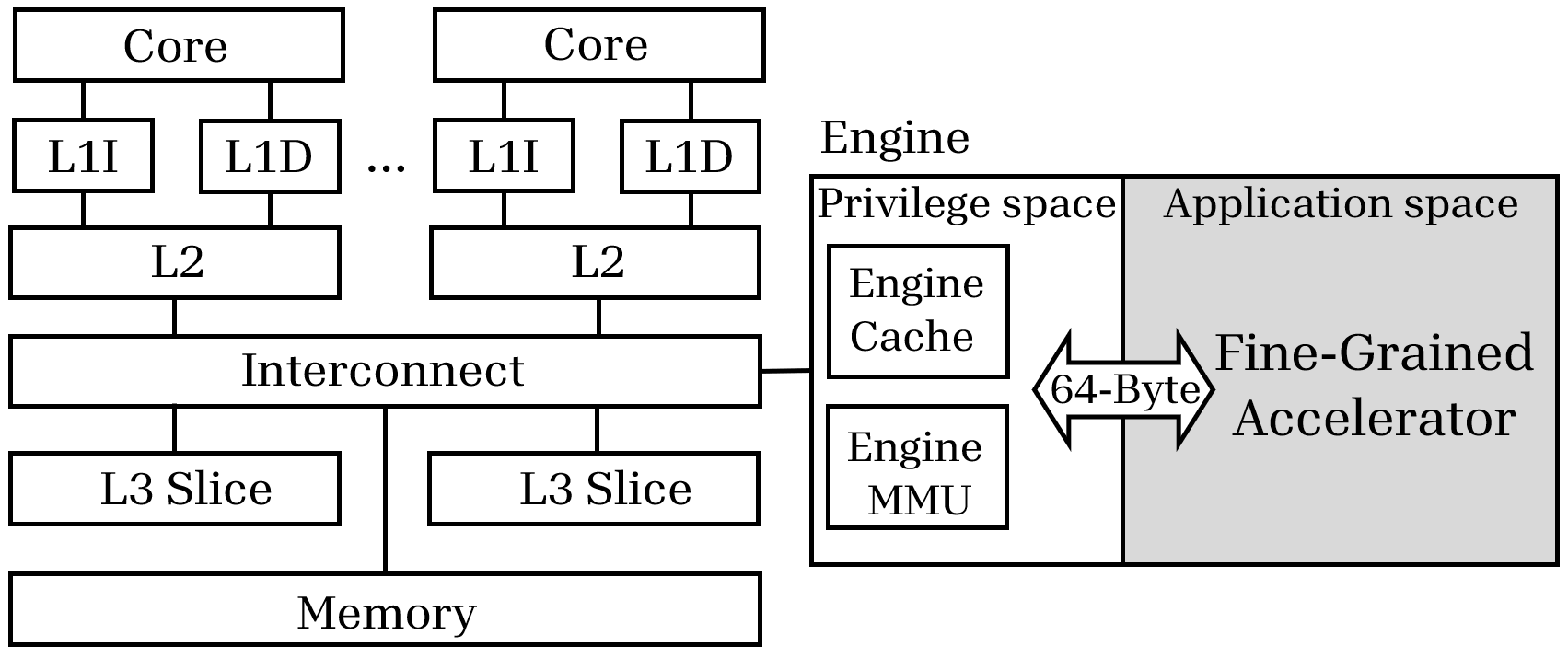}
\caption{High-level system overview of \frameworkname.}
\label{fig:engine_overview}
\squeezeup
\squeezeup
\dontsqueezetoomuch
\end{figure}

Such fine-grained interactions, however, present unique system-level challenges. Frequent core-engine communication, complex cache coherency, and address translation overhead can erode performance gains \cite{2021protobuf,2021_SpZip,2018_garbage_collection,2018_HATS,2017_mcache,2016_COMMTM,2012_base_delta_compression,2019_PHI,2004_adaptive_cache-compression,2017_worklist,2018_minnow,2018_eventriggerpf}. Existing frameworks, primarily designed for large, throughput-oriented accelerators, lack the granularity to model these interactions comprehensively, leading to inaccurate evaluations of fine-grained engine designs. For example, the effective speedup reduces as the latency and memory access for the compute unit increases \cite{2017_logca}. This creates a pressing need for simulation tools that provide holistic system-level modeling, capturing both hardware and software overheads.

To address these challenges, we introduce \frameworkname, an open-source\footnote{We plan to release this tool to open-source when this paper is published.\\AMD, the AMD Arrow logo and combinations thereof are trademarks of Advanced Micro Devices, Inc.} simulation framework built on the gem5 \cite{binkert2011gem5,lowe2020gem5} platform to evaluate fine-grained engines in high-performance systems. Our proposed framework integrates a detailed hardware stack, as illustrated in Figure \ref{fig:engine_overview}. The framework includes a Linux-based software environment running on top of a detailed cache model, along with task-offloading interfaces and address translation mechanisms designed specifically for fine-grained tasks. These features enable the framework to capture complex system-level interactions and provide accurate performance insights.

\frameworkname \space offers
a framework for evaluating an engine performance in a full-system simulation, in which the engine is integrated to a detailed cache system by default.
While gem5 offers great configurability, Choreographer alleviates designers from developing and configuring the many facets of integrating a task-specific engine into a system, such as communication protocols and address translation infrastructure.
In addition, \frameworkname \space delivers both a holistic overview and detailed insights into the system with the presence of the accelerator.
These features enable accelerator designers to study the feasibility and performance of their target accelerator.

We demonstrate the capabilities of \frameworkname \space through two case studies: a data-aware prefetcher for graph analytics, and a quicksort accelerator. These studies illustrate the framework’s ability to model fine-grained performance dynamics and optimize engine designs, emphasizing its value in advancing domain-specific computing.

The key contributions of this work are listed below:
\begin{itemize}
    \item A comprehensive framework, called \frameworkname, for high performance modern cache hierarchy,
    \item A generic communication protocol  
    without requiring instruction set architecture (ISA) change, 
    \item Case studies demonstrating the benefits of the proposed \frameworkname \space framework. 
\end{itemize}

\section{Background}
\vspace*{-\baselineskip}
\label{sec:background}

\begin{center}
\begin{table}[t]
 \caption{Comparisons of NDC Frameworks and \frameworkname}.
 \resizebox{\columnwidth}{!}{
\begin{tabular}{ wc{6em} wc{6em} wc{6em} wc{6em} wc{6em}  } 
  & gem5-Aladdin & gem5-SALAM & NDPmulator & \frameworkname \\ 
  \hline
  \makecell{Simulation \\ Methodology} & Trace-based & \makecell{Execution-based} &  \makecell{Execution-based} & \makecell{Execution-based} \\
 \hline
  \makecell{Accelerator \\ Address \\ Translation \\ Support} & \makecell{Customized\\page tables\\generated from \\ traces} & \makecell{No official\\support for\\ OS-managed \\ pages} & \makecell{Linear translation\\for specific memory \\ region available \\ to accelerator} & \makecell{OS-managed \\ page tables; \\ engine-attached \\ TLB and PTWs} \\
 \hline
 \makecell {Shared L3} & inclusive & inclusive & inclusive & victim \\
 \hline
 \makecell {NoC Modelling} & not-included & not-included & not-included & Detailed Mesh NoC \\
 \hline
  \makecell{Simulation \\ Mode} & gem5 SE & \makecell{gem5 FS \\ bare-metal} & gem5 SE/FS & \makecell{gem5 FS \\ Linux}\\
 \hline
 \makecell{Primary \\ Use Case} & \makecell{General NDC} & \makecell{General NDC} & \makecell{General NDC} & \makecell{Latency-sensitive,\\fine-grained tasks}\\
 \hline
\end{tabular}
}
\label{table:framework_comparisons}
\squeezeup
\squeezeup
\end{table}
\end{center}

This section provides an overview of the key challenges associated with fine-grained accelerators in modern computing systems, motivating the \frameworkname{} framework.
\subsection{Characteristics of Fine-grained Tasks}

Fine-grained tasks are latency-sensitive operations that benefit from accelerators situated within the cache hierarchy. Examples include read-modify-write operations and small-scale serialization tasks. Because the latency overhead of offloading these tasks can be comparable to their execution latency, accurate performance modeling is essential. A common design requirement for handling these tasks is to avoid high-overhead data synchronization mechanisms, such as cache flushing and direct memory access (DMA), by leveraging existing cache coherence protocols~\cite{2024Leviathan,2021protobuf}.

\squeezeup
\dontsqueezetoomuch
\subsection{Importance of Cache Model and Address Translation for Fine-grained Tasks}

\begin{figure}[t]
\centering
\includegraphics[width=0.4\textwidth]{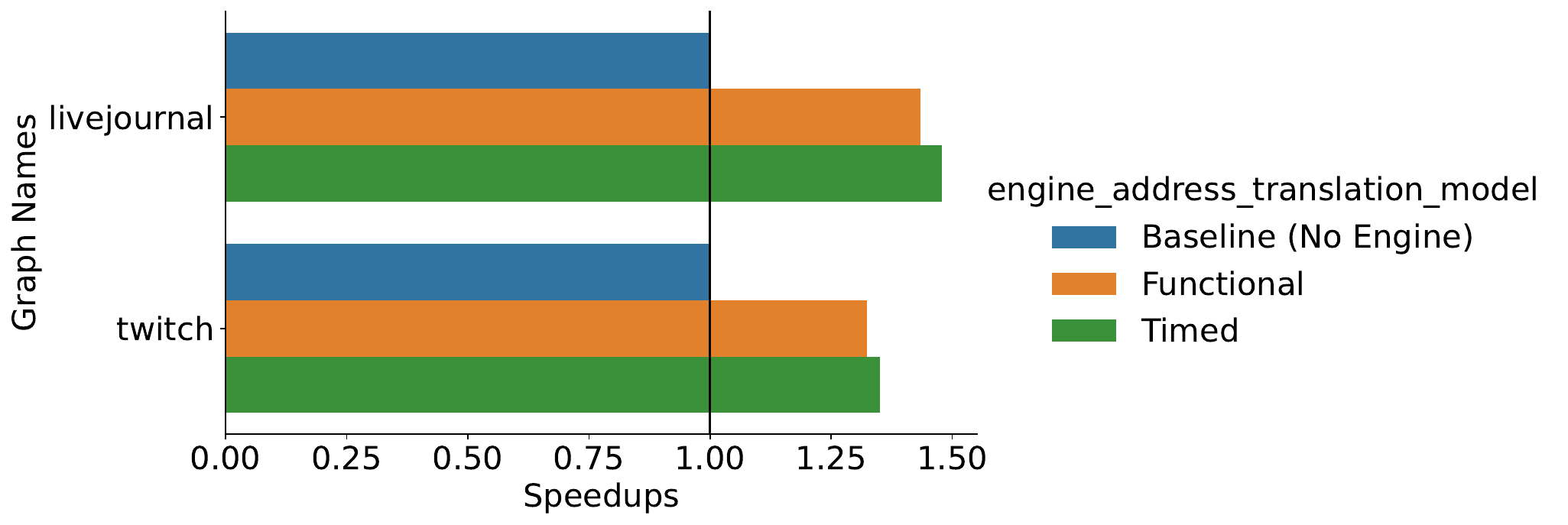}

\caption{
Address translation in an accelerator can affect overall system performance in an unexpected way. 
Because functional translation provides translations instantly, one would expect timed translation to yield lower speedups.
But functional translations underestimate prefetcher performances by up to 4.3\%.
}
\squeezeup
\squeezeup
\dontsqueezetoomuch
\label{fig:translation_effect}
\end{figure}

Accurate cache modeling is critical for evaluating performance of fine-grained tasks. Non-inclusive and exclusive cache policies, commonly employed in modern CPUs, improve effective associativity and capacity, significantly reducing cache misses \cite{2007_nonincl_cache, 1994_exclusive_cache}. The performance difference was
shown to be up to 20\% \cite{2010_achieve_noninc_perf}. These designs also introduce complex trade-offs in latency and performance as core counts increase \cite{2021_l3interconnect}. Without modeling advanced network-on-chip (NoC) effects, which account for up to 75\% of cache miss latency \cite{2010_onchip_interconnect}, frameworks risk underestimating overheads.

Address translation adds further complexity; accelerators often work on OS-managed virtual memory regions, requiring virtual-to-physical address mappings. This process involves up to four additional memory accesses for page table lookups, introducing significant latency. Figure 2 demonstrates the impact of address translation on the system performance, showing how underestimating translation latency can unexpectedly result in pessimistic speedup predictions. For example, incorporating accurate translation modeling yielded up to a 4.3\% speedup improvement in certain configurations, emphasizing its importance in realistic evaluations.

\subsection{Interface to custom accelerators}
Instruction-based schemes are commonly employed for task offloading, particularly in the context of in-core accelerators~\cite{2021protobuf,2021_SpZip,2018_garbage_collection,2018_HATS,2017_mcache,2016_COMMTM}. These approaches require instruction set architecture (ISA) extensions~\cite{guide2011intel}, as well as modifications to the microarchitecture and compiler. They also necessitate static scheduling or frequent status checks to manage limited accelerator resources, complicating compatibility and performance in systems that utilize speculative execution and instruction reordering.
Due to these complexities, very few, if any, of the proposed accelerators have made their way into commercial CPUs. Nevertheless, this approach remains highly effective and feasible in high-value domains, such as multimedia extensions (MMX)~\cite{1997_MMX} and Intel® advanced matrix extensions (AMX)~\cite{2021_AMX}.

Memory-mapped I/O (MMIO) is a method for accessing peripheral device registers using memory addresses. One way to implement this is by employing an uncacheable (UC) memory attribute on the target memory addresses. Data read from or written to a UC memory region is not stored in the CPU's cache, which means that multiple reads or writes to an address in that region will be serialized. Modern CPUs support this method, eliminating the need for compiler updates. Additionally, due to the large address space supported by contemporary systems, MMIO can expose a substantial number of accelerator registers. However, a drawback of this approach is that only the operating system can set the UC attribute, necessitating modifications to the application software stack.

\subsection{Integrating custom accelerators to CPUs}
An accelerator needs to be integrated where it can acquire necessary data from the CPUs. For example, an accelerator that improves branch prediction accuracy must be inside the core, where branch history is accessible with reasonable latency. Accelerators that need access to data movement can be integrated along the data flow, ranging from in the core (L1, L2), on custom interconnects (e.g., Dagger~\cite{2021_dagger}), or on the PCI bus (e.g., RpcNic~\cite{2025_rpcnic}). Each placement offers different latency, bandwidth, and coherency characteristics that impact workload suitability ~\cite{2017_logca}.

\subsection{Prior Frameworks for Fine-grained Tasks evaluations}

Table \ref{table:framework_comparisons} provides a comparison of open-source frameworks for evaluating NDC and fine-grained tasks. Frameworks such as gem5-Aladdin \cite{shao2016gem5aladdin} and gem5-SALAM \cite{rogers2020gem5salam} provide insights into power and area for accelerator designs but lack system-level integration, including support for OS-managed page tables and NoC modeling. While NDPmulator partially addresses these gaps with device driver integration, it still lacks sufficient NoC and cache modeling capabilities \cite{vieira2024ndpmulator}. These shortcomings highlight the need for a comprehensive framework, such as \frameworkname, that captures hardware and software interactions to provide realistic performance evaluations of custom engines for fine-grained tasks.

\section{\frameworkname ~Framework}
\label{sec:framework}

\frameworkname \space is a simulation framework built on top of gem5, a full-system cycle-level simulator \cite{binkert2011gem5,lowe2020gem5}.
The framework leverages gem5's out-of-order CPU model, and detailed implementation of cache coherence protocols to enable full-system evaluation of the target accelerator.
The framework also follows a similar approach of gem5-resources \cite{bruce2021enabling} for building reproducible full-system simulation artifacts.

Enabled by gem5, \frameworkname \space captures critical system-level interactions between hardware and software, modeling a multi-core CPU with an accelerator engine positioned near the shared cache (Figure \ref{fig:engine_overview}). The core can initiate communication with the engine. While the engine cannot initiate communication with the CPU core directly, it can access the cache hierarchy via a 64-byte wide memory interface. Hence, shared memory techniques enable such interactions if needed.

The engine comprises two separate sections: application space, which implements the accelerator in virtual address space, and privilege space, which manages system-level operations beyond basic memory access. This separation ensures system stability by preventing a malicious accelerator from disrupting the system.

\subsection{Software and Hardware Interfaces} \label{sec:offloading}

Fine-grained tasks encompass a wide range of domains, each individually small but collectively accounting for a significant portion of CPU workloads. Introducing an ISA extension for every one of these smaller domains incurs considerable design and maintenance costs, which can place unnecessary burdens on other customers with different workloads. Consequently, \frameworkname \space offers memory-mapped I/O (MMIO) as the default method for interfacing with the engine. However, due to the flexibility of gem5, an instruction-based method can also be implemented if desired.

\begin{figure}[t]
\centering
\includegraphics[width=0.5\textwidth]{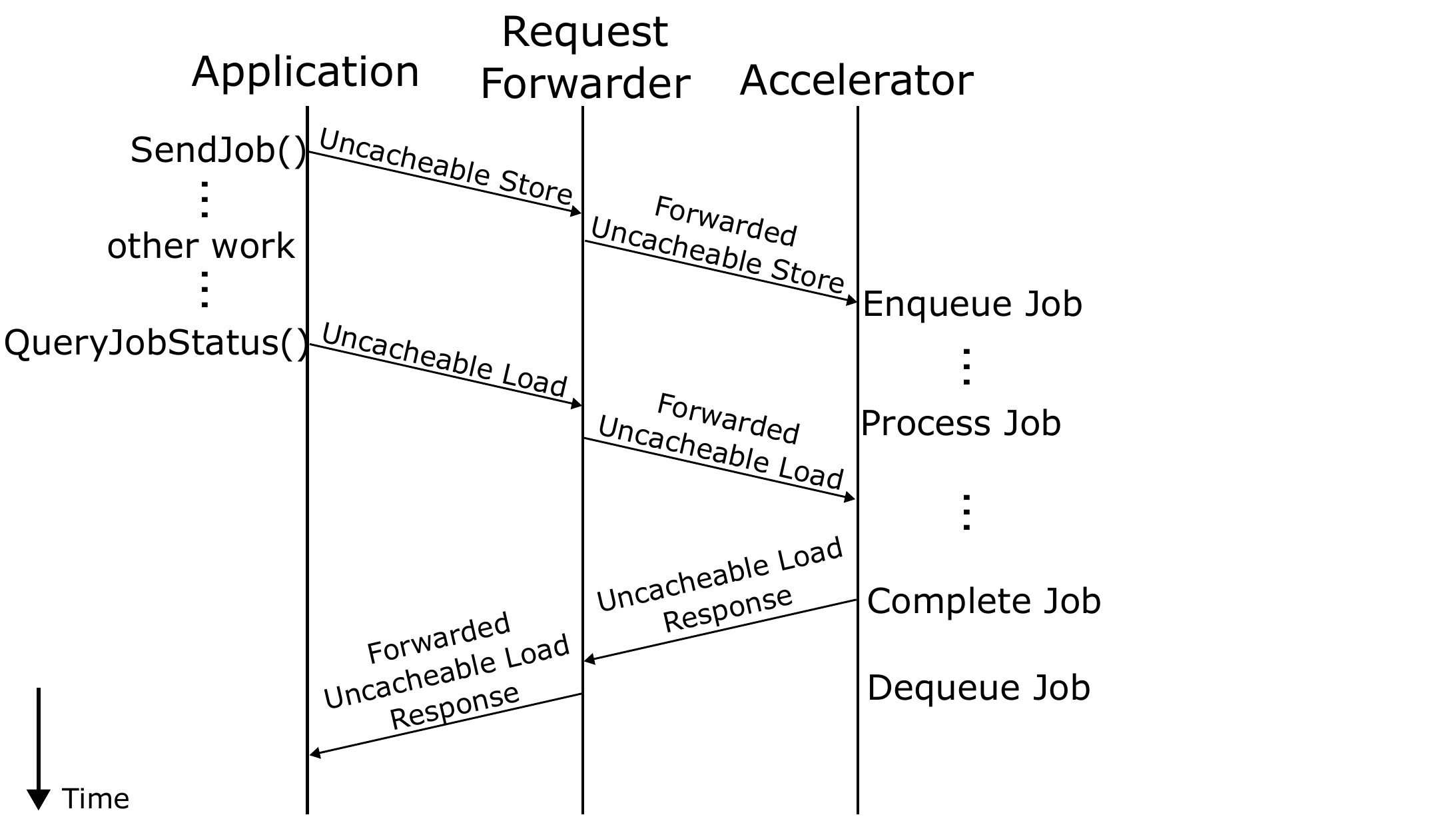}
\squeezeup
\caption{
Our framework task offloading flow. Prior to the first task offloading, the application asks the driver to allocate an uncacheable page, the physical address range of which is sent to the accelerator configuration channel (e.g., MSRs for X86-64 ISA). A task is sent to the accelerator from the core/application via an uncacheable store request to the uncacheable memory region known by the accelerator. The task status can be queried using an uncacheable load request to the accelerator to the same memory region. As gem5's implementation of the CHI coherence protocol which does not support uncacheable memory requests, we introduce a forwarder object in the simulator which forwards uncacheable requests/responses in a specific memory region between the core/application and the accelerator.
}
\label{fig:OffloadingFlow}
\squeezeup
\squeezeup
\end{figure}

\frameworkname{} simplifies task offloading by utilizing uncacheable memory requests managed through the core's load/store queues.
Figure \ref{fig:OffloadingFlow} illustrates the task offloading flow, where the uncacheable memory region is created at runtime by the driver and library (detailed in Sections \ref{sec:driver} and \ref{sec:library}). 
While instruction-based schemes are ISA specific, our scheme is ISA-agnostic. Uncacheable page addresses are sent to the Request Forwarder object so that it can forward requests targeting those addresses to the accelerator with a fixed latency.

Uncacheable requests must be executed in-order with respects to each other; however, non-conflicting loads can bypass an uncacheable request\cite{devices2024amd64}. Our approach uses existing infrastructure, such as using miss status/handling registers, to track offloaded tasks.
As shown in Figure \ref{fig:OffloadingFlow}, we use an uncacheable store for offloading a task, and an uncacheable load to get the status of the task.
Issuing an uncacheable load for querying the task status is optional and application dependent. 
The engine and the application must work in coordination during the offloading process.
When querying the task status, the uncacheable load instruction must be scheduled immediately after the corresponding store instruction to avoid potential deadlocks when the accelerator job queue is full.
The accelerator has a separate communication queue for each core.

\subsection{Accelerator Interface with the Memory System}
A fine-grained task that operates on data can be executed at various levels within the cache hierarchy, ranging from the core's private L1 and L2 caches to the shared last level cache (LLC) and down to the memory controllers. Locating the engine closer to the core would necessitate significant design and verification efforts due to the highly optimized nature of the core. Conversely, placing the engine outside of the core system-on-chip (SoC) would introduce excessive communication latency, making it less responsive to the core. Therefore, we have positioned the engine near the LLC as shown in Figure~\ref{fig:engine_overview}, where it can natively participate in cache coherency while remaining within a short reach of the core.

In addition to handling commands from the cores, most accelerators require access to data in memory. \frameworkname ~provides 64-byte wide load and store interfaces that operate exclusively in virtual address space. Address translation is managed by the privilege space of the engine as described in the next section. Since the need for a store buffer varies by accelerators, Choreographer does not include one by default, leaving its implementation to the specific accelerator design. An example of such an implementation is discussed in Section \ref{sec:quicksort}. 

\subsection{Address Translation}
\label{sec:addr_translation}
As the accelerator operates exclusively in virtual address space, the engine's memory management unit (MMU) ensures proper handling of all pages sizes. The MMU has the same structures as that of the cores, which includes TLBs and PTWs. If the accelerator accesses a page that is not present in the physical memory but is present in the virtual address space, the request is dropped without triggering a page fault. The accelerator will be notified of such fault and can issue the request again later. A fine-grained accelerator operates in a fine interaction to the cores. Therefore, this situation would rarely occur and is expected to be resolved soon after.

\subsection{The Cache Model} \label{sec:CacheModel}

\begin{figure*}[t]
\centering
\includegraphics[width=\textwidth]{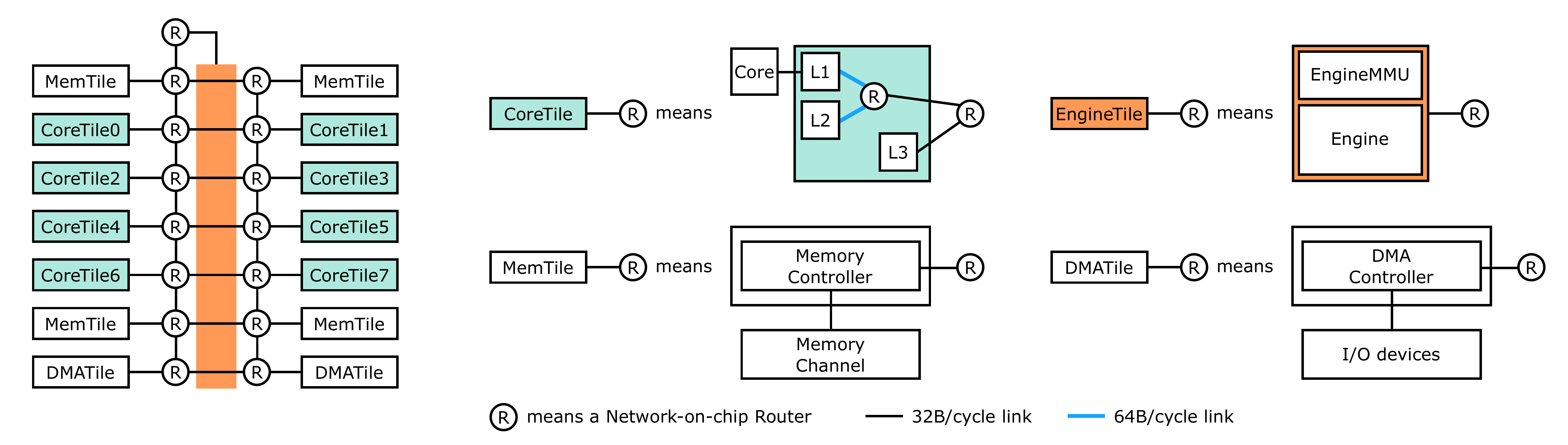}
\squeezeup
\squeezeup
\dontsqueezetoomuch
\dontsqueezetoomuch
\dontsqueezetoomuch
\caption{Illustration of our detailed cache model included in our framework.}
\label{fig:DetailedCache}
\squeezeup
\squeezeup
\dontsqueezetoomuch
\end{figure*}

\textbf{Setup.}
Figure \ref{fig:DetailedCache} illustrates the one core complex die (CCD) setup corresponding to the Mesh-8cores in our prebuilt cache configuration. The model is built on gem5's RUBY implementation of the CHI cache coherence protocol \cite{amba_chi} and uses a tile-based abstraction, which including the following:
\begin{itemize}
    \item CoreTile: Contains one out-of-order core, split private L1 caches, a unified L2 cache private to the core, and a slice of the shared L3 victim cache.
    \item L3OnlyTile: Includes only a slice of the shared L3 victim cache to simplify simulation when only some cores are used.
    \item EngineTile: Houses the accelerator engine, its private cache, and a memory management unit.
    \item MemTile: Contains a memory controller handling transactions for a specific memory channel.
    \item DMATile: Includes a DMA controller.
\end{itemize}

These tiles are arranged in a mesh topology, with the engine placed at one end. All cache levels operate at the same clock frequency as the core, while the engine operates at a lower clock frequency. For the rest of the paper, we refer to "cycle" as the core's cycle unless stated otherwise. Table \ref{table:SystemConfig} lists default system configurations, in which the cache parameters are configured to match those of "Zen 5" \cite{2024zen5hotchips}.

\textbf {Configurability.}
The cache model provides Python classes for each tile abstraction, allowing framework users to rearrange the tiles as needed. The framework also includes a MeshDescription class that simplifies the setup of mesh-like cache topologies, enabling flexible configurations to suit various design requirements.

\subsection{Accelerator Driver} \label{sec:driver}
The driver is a privileged software component essential for enabling user applications to utilize an accelerator. It performs the following operations:

\begin{itemize}
    \item Detects engine availability and either initializes it or defaults to a software implementation if the engine is unavailable. This allows the same application binary to run on systems with or without the engine. 
    \item Transfers initialization data from the application to the accelerator.
    \item Allocates an uncacheable page for the application and sends corresponding physical addresses to the accelerator. The application sends commands and other information via this page (Section \ref{sec:offloading}).    
\end{itemize}

\subsection{Domain-specific Library} \label{sec:library}

While a driver is sufficient for applications to utilize an accelerator, direct interaction with it can be cumbersome and error-prone. For domains like graph analytics, where similar information is required across applications, a domain-specific library can simplify development by providing a high-level API. Such a library can often integrate seamlessly with existing application frameworks, eliminating the need for extensive application modifications.

As part of \frameworkname, we provide a utility library to streamline application development with accelerators. Additionally, we include an example graph-specific library for GAPBS \cite{beamer2015gap}, which was used in the experiments detailed in Section \ref{sec:prefetcher}.

\subsection{Putting It All Together}
The framework enables integration of the target accelerator into a system with out-of-order cores, an engine integrated with a detailed cache model, and an address translation facility for the engine.
\frameworkname \space also offers a software integration API via the accelerator driver and the domain-specific library, which are responsible for resource management and communication between software and the accelerator.
\section{Experiment Methodology}

We use the \frameworkname \space framework for all experiments. Each workload is evaluated on two systems: the baseline system, where the engine is inactive, and the accelerated system, where the engine is activated during workload execution. Both systems share identical parameters (Table \ref{table:SystemConfig}), except the baseline system does not utilize the engine. An inactive engine does not interact with the rest of the system, and subsequently, does not affect the performance of the baseline system.

\begin{figure}[t]
\centering
\includegraphics[width=0.4\textwidth]{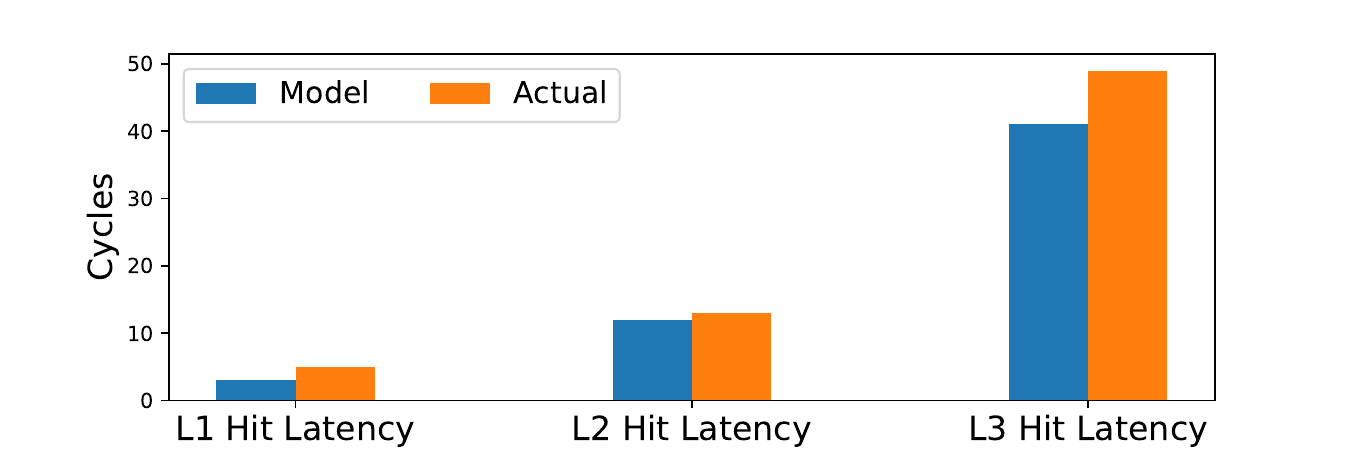}
\caption{Correlation of cache hit latencies to a real system.}
\label{fig:hit_latency}
\squeezeup
\squeezeup
\end{figure}

The experiments utilize the cache topology illustrated in Figure \ref{fig:DetailedCache}, with latencies carefully calibrated to closely align with those of "Zen 4," as depicted in Figure \ref{fig:hit_latency}.
 To ensure accurate and fair comparisons, we utilize process pinning and simulation checkpointing.

\textbf{Process Pinning for single-threaded experiments.}
We keep CoreTile0 and CoreTile1, while other CoreTiles are replaced by L3OnlyTile. Workloads are pinned to CPU 1 (CoreTile1), leaving CPU 0 (CoreTile0) idle for kernel tasks, which reduces scheduling interruptions and minimizes context switches. Hardware characteristics are consistently captured on core 1, and its performance statistics are used for comparison.

\textbf{Simulation Checkpointing.}
Checkpointing saves and restores the system's architectural state. Using kernel virtual machine (KVM), simulations are fast-forwarded to the start of the application, and checkpoints are saved. Both systems are restored from the same checkpoint to ensure identical initial states and workload binaries. Guest software parameters, such as engine activation, are injected during restoration for consistent evaluation.

\begin{center}
\begin{table}[t]
\footnotesize
 \caption{Default System Parameters Used for Evaluation.}
 \squeezeup
 \resizebox{\columnwidth}{!}{
\begin{tabular}{ wc{6em} wc{24em}  } 
 \hline
  \textbf{Cores} & \makecell{2 out-of-order cores @ 4GHz, x86-64/ARM ISA\\ No simultaneous multithreading (no SMT) \\ 20-core-cycle latency for forwarding uncacheable requests to engine}\\
  \hline
  \textbf{Core TLB} & \makecell{private split instruction/data\\1 level, 64 entries, fully associative \\ No process context identifiers (no PCID) support \\ 4-level pages with HugePage support, 1 PTW}\\
  \hline
  \textbf{L1 Cache} & \makecell{private split instruction/data caches\\Capacity: 32KiB for L1I and 48KiB for L1D \\ 8-way set associative for L1I and 12-way set associative for L1D \\ L1I has a stride prefetcher} \\
  \hline
  \textbf{L2 Cache} & \makecell{private unified cache, 1MiB capacity, 16-way set associative} \\
  \hline
  \textbf{L3 Cache} & \makecell{victim cache, shared among all cores\\ 32MiB total capacity 
(4MiB per CoreTile/L3OnlyTile) \\ 16-way set associative} \\
  \hline
  \textbf{Memory} & \makecell{3GiB, 4-channel DDR4 2400 \\ 19.2GiB/s theoretical maximum bandwidth} \\
  \hline
  \textbf{Engine} & \makecell{
    1 engine @ 1GHz clock frequency\\
    512-KiB private data cache, 8-way set associative
  } \\
  \hline
  \textbf{Engine TLB} & \makecell{1-level 1-entry fully-associative data TLB, no PCID support \\ 4-level pages with HugePage support, 1 PTW 
  } \\
  \hline
  \textbf{Operating System} & \makecell{
    Ubuntu Server 22.04.5 LTS\\
    Linux Kernel v5.15.141, Transparent Hugepage Enabled
  } \\
  \hline
  \textbf{Guest Compiler} & \makecell{
    GCC 11.4.0
  } \\
  \hline
  \textbf{gem5} & \makecell{
    v24.0.0.1
  } \\
  \hline
\end{tabular}
}
\label{table:SystemConfig}
\squeezeup
\squeezeup
\end{table}
\end{center}

\squeezeup
\squeezeup
\squeezeup

\section{Case Studies}
\label{sec:cases}
In this section, we present two case studies that exemplify data-triggered and small-offloaded tasks, which represent the target use cases of  \frameworkname.
The first case study examines a data-aware prefetcher, where the core frequently sends prefetch hints to the prefetcher.
The second case study builds a quicksort accelerator, where the core offloads the entire task of sorting an array to the accelerator.
The different task granularity and offloading strategies show the versatility of the framework.
We also highlight the benefits of full-system simulations, in which the engine works in coordination with the workload and the operating system for task offloading and address translation.
\frameworkname \space provides features and configurations beyond our case studies. For those features not listed here, we provide examples along with the \frameworkname \space codebase.

\subsection{Case Study I: Data-aware Prefetcher}
\label{sec:prefetcher}

\begin{figure}[t]
\centering
\includegraphics[width=0.48\textwidth]{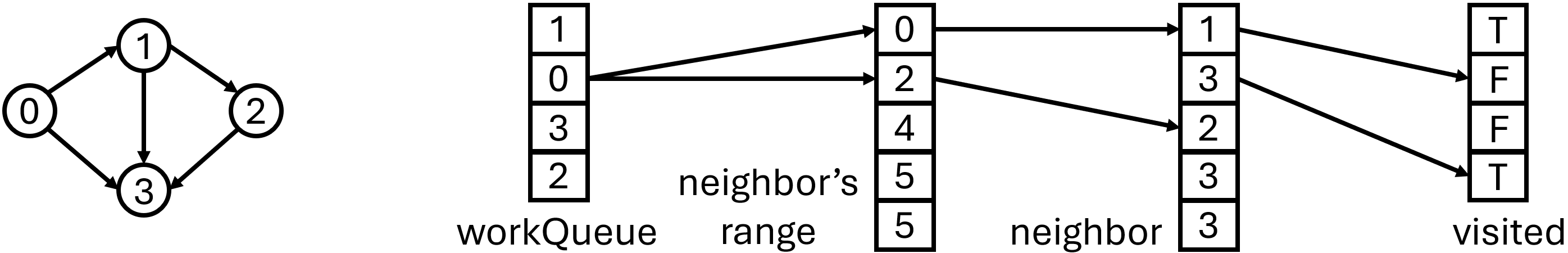}
\caption{A DIG for BFS on a simple graph.}
\label{fig:simpleBFSgraph}
\squeezeup
\squeezeup
\end{figure}

In this study, we implement a Prodigy-like prefetcher using the data indirection graph (DIG) to drive the prefetcher \cite{talati2021prodigy}.
The DIG contains information on the chain of indirect accesses between data structures.
Figure \ref{fig:simpleBFSgraph} shows the DIG of GAPBS's implementation of the top-down breadth-first search (BFS) algorithm, which contains four levels of indirections.
The first access is to a work queue, which is used to pop a node to process.
This node is used to index the neighbor's range array, which points to the list of the node's neighbors. Finally, each neighbor entry indexes the visited array that keeps track of whether a node has been visited earlier.
This chain of indirections repeats for every node in the work queue.

Our prefetcher starts prefetching when the node on top of work queue is accessed.
The access triggers a series of memory requests to prefetch the data of the indirection tree needed for the node that is $K$ nodes ahead of the accessed node.
Unlike Prodigy, which prefetches useful cache blocks to the core's private cache, our prefetcher loads data blocks to the private engine cache.
Even though our prefetcher does not prefetch data directly to the core's private cache, we expect the prefetcher to fetch the blocks to the cache hierarchy before the core consumes data in the prefetched blocks, reducing the memory access latency.
From the core's perspective, the latency of accessing its private cache is much lower than that of accessing other private caches or the shared L3.
Therefore, we expect our prefetcher to result in smaller speedups compared to Prodigy.

We use GAPBS's implementation of top-down BFS, without prefetch hints, as the baseline.
In the implementation of top-down BFS loop with prefetch hints, for each node $u$ that is being visited, the core sends a message to the engine to hint prefetching at distance $K$ if there are at least $K$ nodes ahead of $u$ in the work queue, for a fixed number $K$.
The hint is sent to the engine via a store to the uncacheable page allocated at the beginning of the workload.
For each visited node, the overhead of sending the hint consists of nine instructions including a load with high temporal locality, a branch depending on the load to check for the queue size, and an uncacheable store sending the hint.
A more optimized implementation can optimize away the extra load.

We evaluate our prefetcher in both single-threaded and multi-threaded settings. First, we examine the single-threaded scenario to gain insights with minimal noise that typically accompanies multi-threaded environments. Subsequently, we validate the key observation using the multi-threaded setting.

\subsubsection{Prefetching for Single-Threaded Applications}
For benchmarking, we use different sparse graphs with different sizes and degrees to evaluate the effectiveness of the prefetcher in different settings.
The degree of a node in an undirected graph is the number of its neighbors, and the degree of a graph is the average number of neighbors of each node.
We use three synthetic Kronecker undirected graphs synthesized using GAPBS with characteristics shown in Table \ref{table:graph_spec}, as well as LiveJournal \cite{yang2012defining} and Twitch \cite{rozemberczki2021twitch} graphs.
We observe that synthetic graphs have a distinct distribution of the number of neighbors compared to real graphs.
As a result, the real graphs are not directly comparable to the synthetic graphs even if they share similar degrees.
The BFS algorithm is modified to run twice for each experiment. The first run is to warm up the cache. The results are the measurement of the second run, which uses a different starting node.
We choose $K \in \left\{1, 2, 4, 8, 16, 32\right\}$ in our experiments.
Since we use simulation checkpointing, and since the single-threaded version of GAPBS is deterministic, the nodes are visited in the same order for each graph across all experiments.

\begin{figure}[t]
\centering
\includegraphics[width=0.5\textwidth]{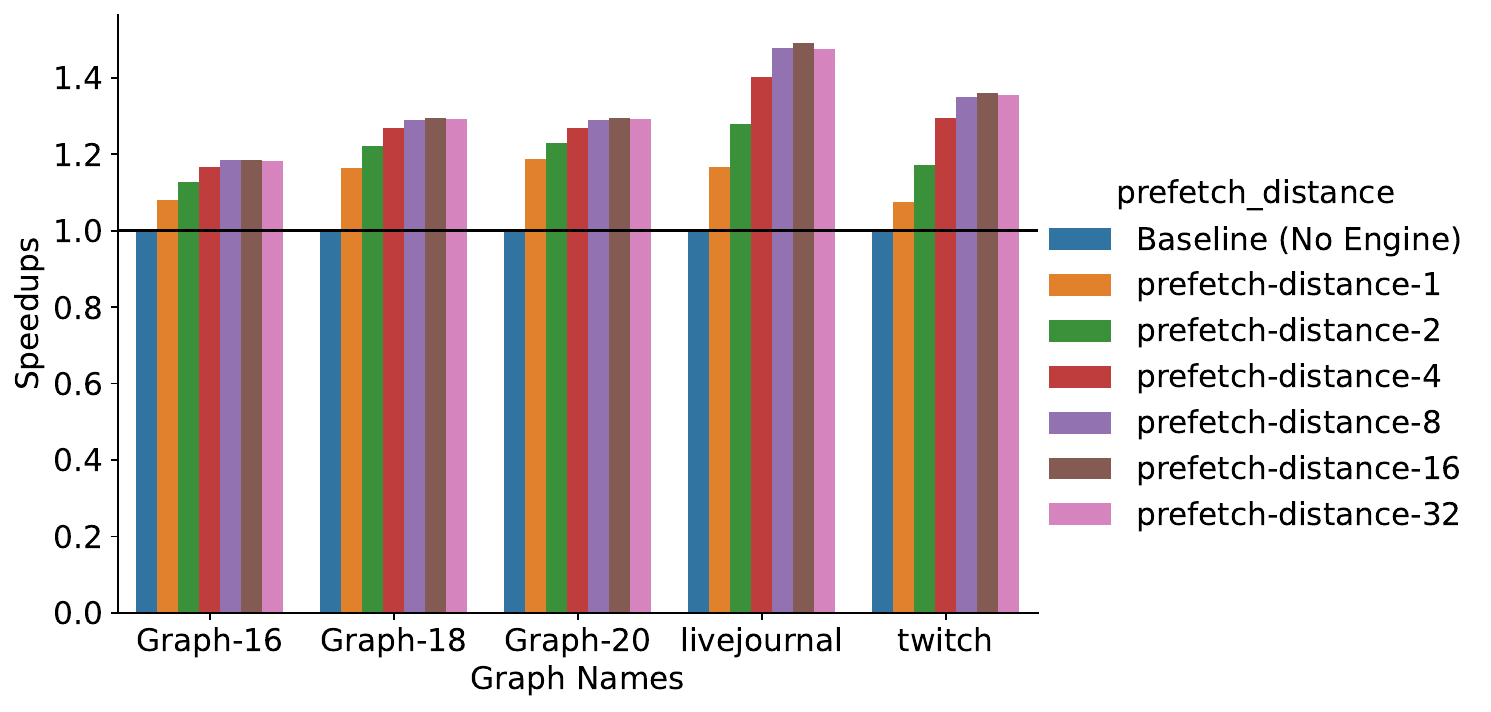}
\squeezeup
\squeezeup
\caption{Speedup of data-aware prefetcher of various distances over the baseline system for different graph sizes. Note that the engine TLB has a capacity of 1 and engine cache has capacity of 512KiB for this experiment.}
\label{fig:prefetcher_speedup}
\squeezeup
\end{figure}

\begin{center}
\begin{table}[t]
\footnotesize
\centering
\caption{Graphs Used in Data-aware Prefetcher Evaluation.}
\begin{tabular}{ | c | c | c | c | c | c |}
\hline
\makecell{Graph \\Name} & \#Nodes & \#Edges & Degrees & \makecell{Memory \\ Usage \\ (MiB)} & \makecell{\#Nodes \\ with \\ degrees \\ $>$ 128}\\
\hline
Graph-16 & $2^{16}$ & 3.4M & 51.9 & 14.0 & 13\%\\
\hline
Graph-18 & $2^{18}$ & 14.5M & 55.3 & 59.3 & 8\%\\
\hline
Graph-20 & $2^{20}$ & 60.9M & 57.8 & 247.5 & 8\%\\
\hline
LiveJournal & 4.0M & 34.7M & 8.7 & 193.3 & 1\%\\
\hline
Twitch & 168.1K & 6.8M & 40.4 & 28.5 & 7\%\\
\hline
\end{tabular}
\label{table:graph_spec}

\squeezeup
\squeezeup
\end{table}
\end{center}
\squeezeup
\squeezeup
\dontsqueezetoomuch
\textbf{Observation 1. The magnitude of speedups obtained by our prefetchers is highly dependent on the size of the graph and the portion of high degree nodes.}
Figure \ref{fig:prefetcher_speedup} shows that, using the default engine configuration, we obtain speedups over the baseline system ranging between 1.08x and 1.49x across all prefetching distances.
The result matches our expectation that the speedup peaks at a certain prefetch distance before slightly decreasing at larger distances.
Our prefetcher does not change the content of the core's private caches, and at high prefetch distances, the speedup slightly decreases.
This is different from private cache prefetchers, which can negatively impact performance at high prefetch distances since the prefetched cache blocks might evict useful data the core needs due to limited private cache capacity.

We observe that the peak performance of the prefetcher for each graph is highly dependent on the portion of high degree nodes in the graph.
We define nodes with more than 128 neighbors as high degree nodes.
Processing a high degree node induces a burst of memory requests to enumerate through its neighbors scattered around memory.
Hence, the ratio of high degree nodes indicates the frequency of the memory system being strained due to a burst of memory accesses.
From Figure \ref{fig:prefetcher_speedup}, we see an inverse correlation between the portion of nodes with high degrees and the peak speedups we obtain for each graph.
Among the benchmarked graphs, Graph-16 has the lowest peak speedup at 1.19x and the highest ratio of high degree nodes at 13\%.
In contrast, the LiveJournal graph has the highest peak speedup at 1.49x and the lowest ratio of high degree nodes at 1\%.

\begin{figure}[t]
\centering
\includegraphics[width=0.5\textwidth]{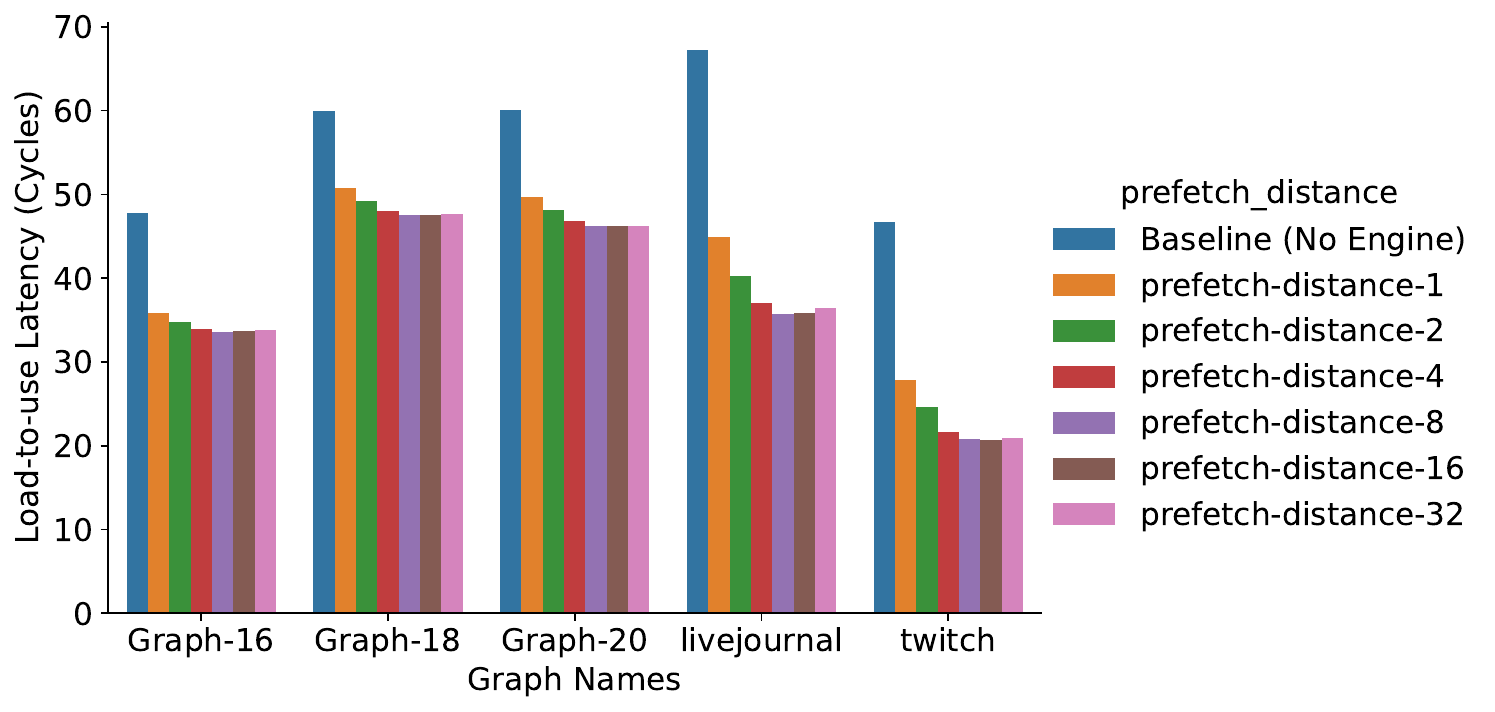}
\squeezeup
\squeezeup
\caption{Latency of load instructions of core 1, which was executing the workload. The lower latency of loads suggests that the cache contains more useful data than that of the baseline system.}
\label{fig:load_use_latency}
\squeezeup
\squeezeup
\end{figure}

To understand the nature of the speedup, we theorize that the load instructions to the graph data structure have lower latency compared to the baseline.
We choose core 1's load-to-use latency as the metric to measure the effectiveness of a prefetcher as the eventual goal of a prefetch is to lower the memory request latency.
Using solely the cache hit rate does not necessarily show the prefetch effectiveness of a prefetcher as a prefetcher can improve the cache miss latency instead of the hit rate.
Note that when the prefetcher is activated, there is an overhead of a high temporal locality load per visited node; however, the overhead is insignificant compared to the number of loads induced by visiting edges.
Figure \ref{fig:load_use_latency} depicts the average load instruction latency for various graph sizes of core 1, the core that performs the algorithm.
The figure shows that the average load latency decreases by up to 25\% and 50\% for the optimal prefetch distance for the synthetic graphs and the real graphs, respectively.
The graph shows a strong inverse correlation between the load-to-use latency and the obtained speedups within a particular graph.
This strongly suggests that the cache hierarchy, with help from our prefetcher, has more useful cache blocks within the cache hierarchy compared to the baseline.

\begin{figure}[t]
\centering
\includegraphics[width=0.5\textwidth]{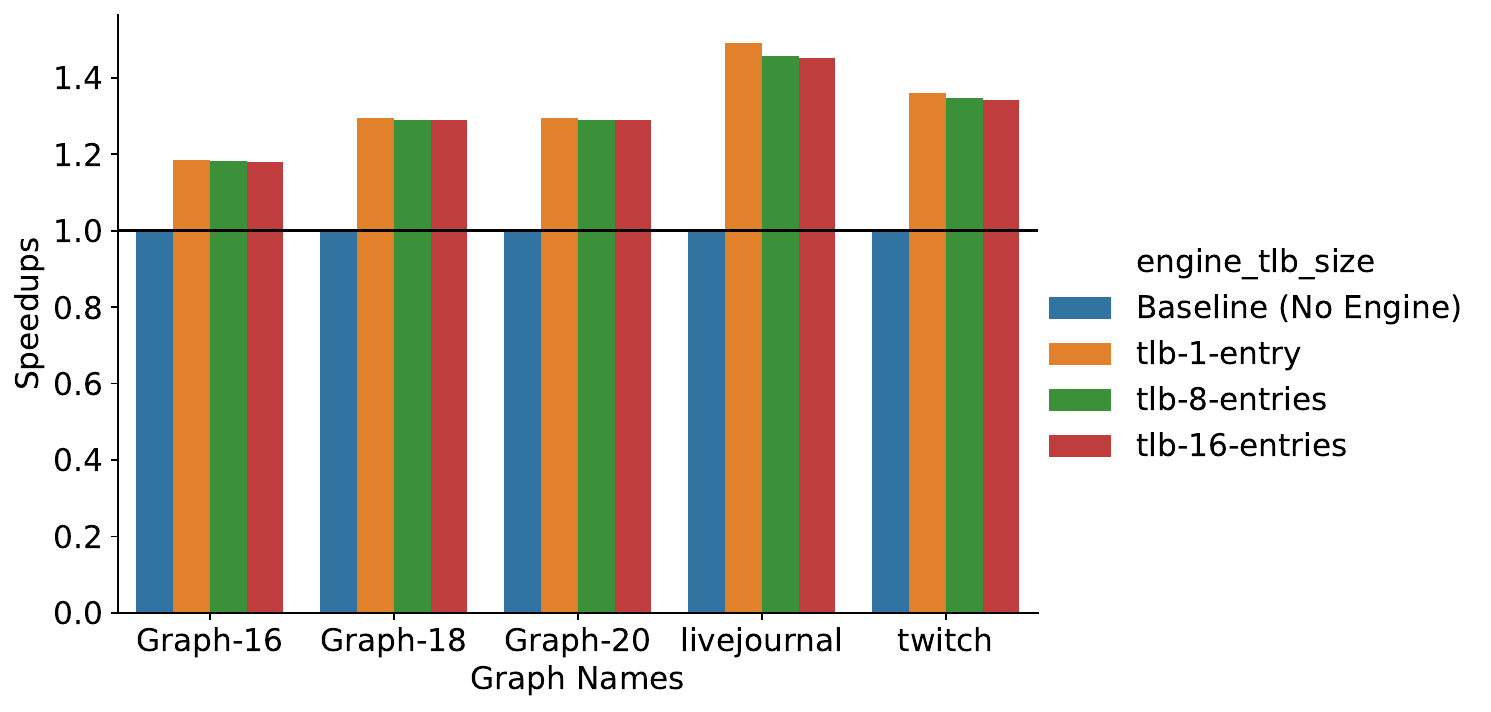}
\squeezeup
\squeezeup
\caption{Comparisons of prefetcher effectiveness with various engine TLB capacities. Note that we fix the prefetch distance to 16 and the engine cache has capacity to 512KiB for this experiment.}
\label{fig:tlb_size_impact}
\squeezeup
\end{figure}

\textbf{Observation 2. Low-capacity engine TLB can improve performance.}
Figure \ref{fig:tlb_size_impact} shows the impact of higher capacity TLB on the prefetcher's performance.
As the capacity of the engine TLB increases, the performance decreases up to 3\% for the LiveJournal graph when increasing the engine TLB capacity from 1 entry to 16 entries.
While the decrease in performance is small and might be due to system's noise, this indicates that the prefetcher might not need a large engine TLB capacity.

\begin{figure}[t]
\centering
\includegraphics[width=0.5\textwidth]{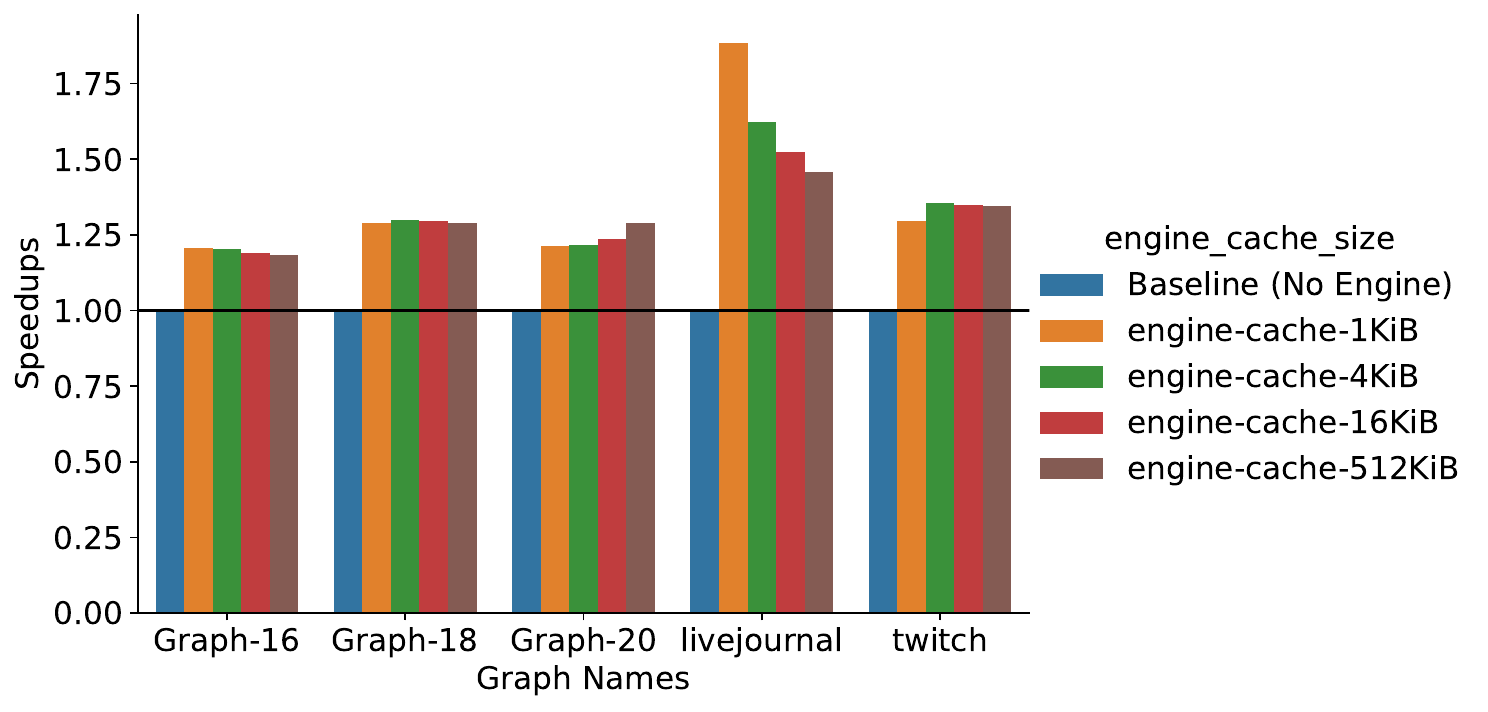}
\squeezeup
\caption{Comparisons of prefetcher effectiveness with various engine cache capacities. Note that we fix the prefetch distance to 16 and engine TLB capacity to 8 for this experiment.}
\label{fig:cache_size_impact}
\squeezeup
\squeezeup
\end{figure}

\textbf{Observation 3. Increasing the engine cache size might negatively impact performance.}
Figure \ref{fig:cache_size_impact} shows the impact of engine cache capacity on the prefetcher's performance.
Notably, a higher engine cache size positively impacts the performance of the Twitch graph, but negatively impacts the performance of the Livejournal graph, which sees the speedup decreases from 1.88x with 1KiB engine cache to 1.46x with 512KiB engine cache.
We believe the smaller engine cache size induces more eviction of useful cache blocks from the engine cache to the shared L3 cache before the core consumes the blocks.
From the core, accessing a block from an L3 cache has a lower latency than accessing a block from the engine private cache.
This effect is observable as \frameworkname \space models the L3 cache as a victim cache.
An inclusive L3 cache does not have this performance difference with different engine cache sizes as all contents in private L2 caches and the engine cache are also in the L3 cache.

\begin{figure}[t]
\centering
\includegraphics[width=0.5\textwidth]{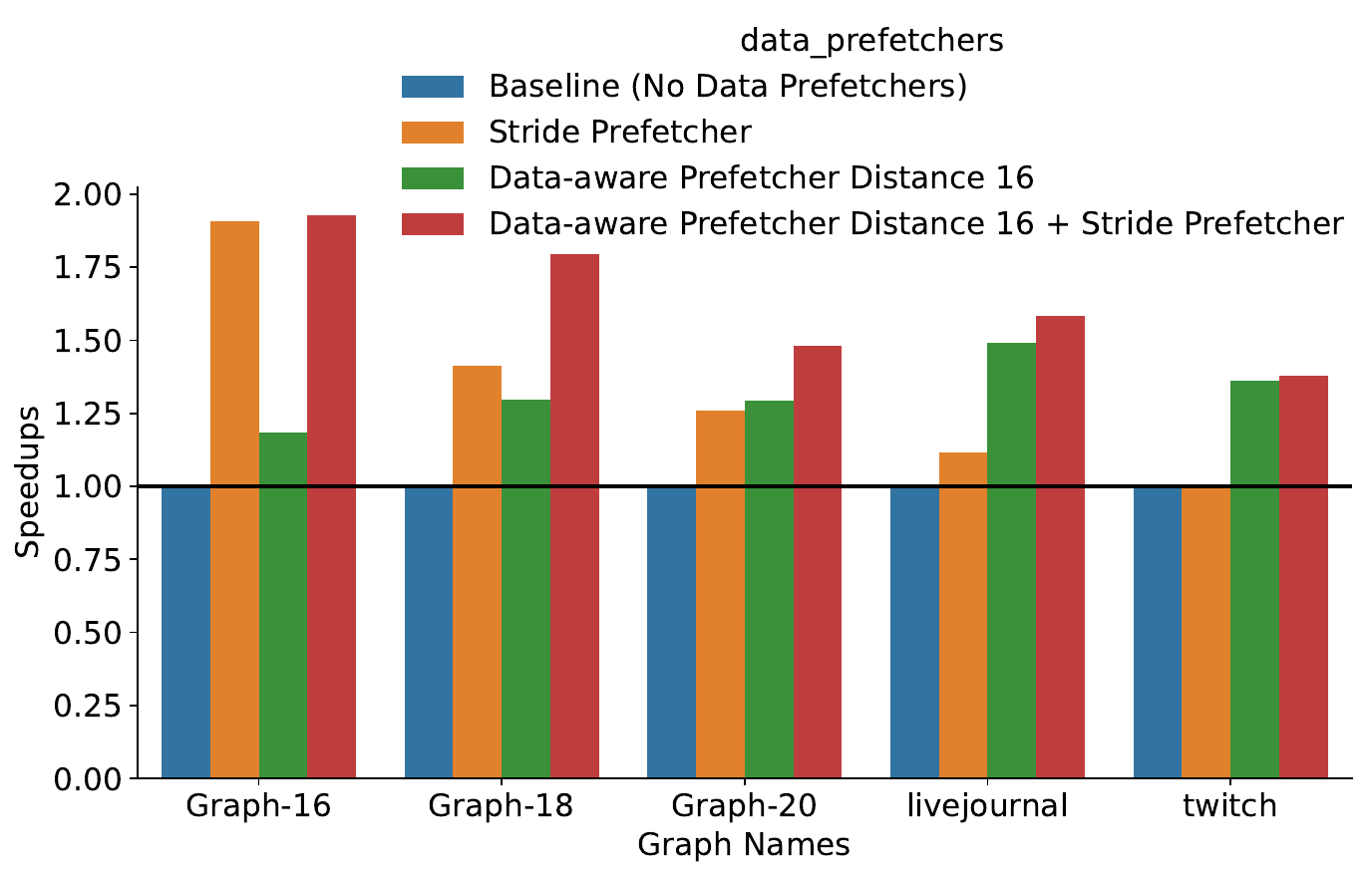}
\squeezeup
\caption{Speedups over the baseline of our prefetcher and core's private cache stride prefetcher. In the setup with "Stride Prefetcher", each of L1D cache and L2 cache has a stride prefetcher.}
\label{fig:compare_with_stride_prefetcher}
\squeezeup
\squeezeup
\end{figure}

\textbf{Observation 4. Our prefetcher should be used with stride prefetchers for better performance.}
Figure \ref{fig:compare_with_stride_prefetcher} shows the performance of our prefetcher with and without the presence of core's private cache stride prefetcher~\cite{baer1991effective}.
For Graph-20, LiveJournal, and Twitch graphs, using our prefetcher yields better performance than only using stride prefetchers.
Notably, for Graph-16, the stride prefetchers obtains 50\% more speedup compared to our prefetcher.
The figure also shows that, using both our prefetcher and stride prefetchers yields better performance than using only one of the prefetchers.
The result suggests that, at a certain prefetch distance, our prefetcher helps bringing data into the cache hierarchy before the stride prefetchers bring the data into the private cache.

\textbf{Takeaways.} The detailed modeling of the cache system and address translation at the engine enable us to observe the counter-intuitive effects of our prefetcher on overall system performance.

\subsubsection{Prefetching for Multi-Threaded Applications.}
\begin{figure}[t]
\centering
\includegraphics[width=0.3\textwidth]{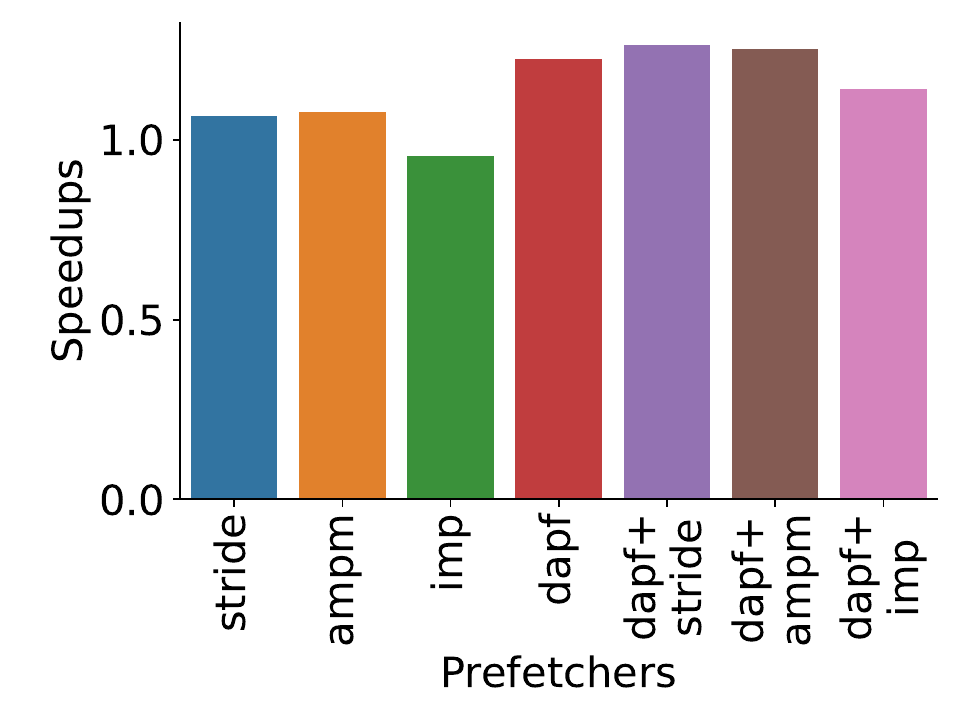}
\caption{Speedups of the LiveJournal graph, utilizing 8 threads. The terms "stride," "ampm," and "imp" indicate scenarios when such prefetcher is implemented in both the L1D and L2 caches. For the data-aware prefetcher (dapf), the prefetch distance is set to 32.}
\label{fig:8t_prefetcher}
\squeezeup
\squeezeup
\end{figure}

The speedups observed in a multi-threaded setting are illustrated in Figure \ref{fig:8t_prefetcher}, based on experiments conducted with eight threads. The findings from Observation 4 remain valid: our prefetcher provides greater speedup compared to the private stride prefetcher used in isolation. However, it is worth noting that the speedups for each combination of prefetchers are lower than their one-thread counterparts, as the memory system's throughput is shared among multiple threads.
Additionally, we compare our prefetcher with Address Map Pattern Matching (AMPM)~\cite{ishii2009access} and Indirect Memory Prefetcher (IMP)~\cite{yu2015imp}, both of which are private cache prefetchers. The results further generalize Observation 4 to these prefetchers. Furthermore, while some private cache prefetchers may degrade performance due to their predictive nature, our prefetcher consistently delivers accurate prefetches and is not susceptible to such errors.


\subsection{Case Study II: Quicksort Accelerator}\label{sec:quicksort}
The quicksort algorithm is an efficient, general-purpose sorting algorithm used in many libraries including the GNU C++ Standard Template Library \cite{Kristo2020_sort}.
The algorithm recursively partitions an array into two contiguous sub-arrays.
An element is moved to one of the sub-arrays depending on whether it is less than or greater than the chosen pivot.
Each sub-array is partitioned using a new pivot. The algorithm ends when every sub-array contains only one element or contains a small number of elements that can be efficiently sorted by other sorting algorithms. 

Our quicksort accelerator implements the algorithm entirely in hardware.
To offload the task of sorting an array to the accelerator, the application sends the virtual address and size of the array using uncacheable requests.
When the information is received, the accelerator issues two load requests: one for a cache line at the beginning and another for the cache line at the end of the array.
The key distinction between our implementation and the software version stems from the granularity of memory access.
In the software version, an element of the array can be read from or written to individually. The read data appear in order even when the data returned from memory are out of order.
The interaction with the memory system is handled implicitly by the underlying hardware mechanisms.
In contrast, an accelerator interacts with memory at the cache line granularity and the returned data can be out of order.
Furthermore, two or more sub-arrays could share the same cache line.
The situation can be efficiently handled using a store buffer that not only provides a load response for any request that hits the store buffer, but also stores a data block until all its sub-arrays have written out their data.

\begin{figure}[t]
\centering
\includegraphics[width=0.35\textwidth]{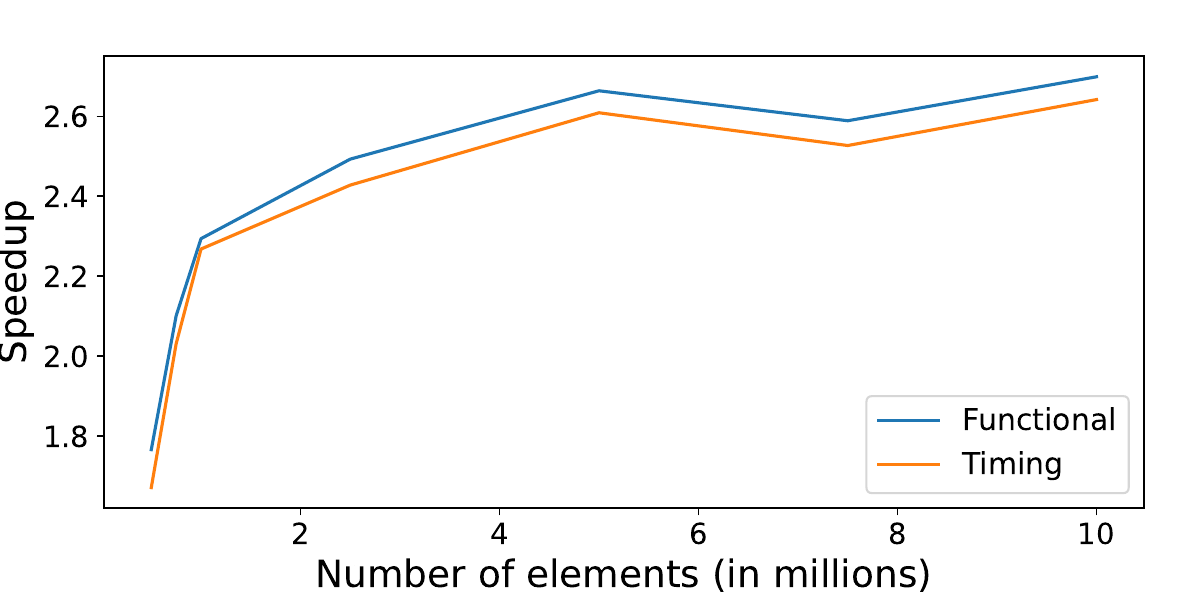}
\caption{The maximum speedup is achieved when the address translation completes in the same cycle (Functional). 
The latency of translation in Timing mode causes speedup reduction.
}
\label{fig:qsort_speedup}
\squeezeup
\end{figure}

The speedup of the accelerator over a software implementation is shown in Figure \ref{fig:qsort_speedup}. Overall, an accelerator provides more than 2x speedup for most array sizes.
The break-even point is well below 500,000 elements. 
Address translation negatively affects speedup as shown in the figure. We evaluated our address translation technique (Section \ref{sec:addr_translation}) on this accelerator. The negative effect of address translation latency is marginal. This is because the accelerator is able to use most of the cache line in a page. Thus, the translation latency is amortized over a large number of usages. 
The results highlights the fact that some accelerators do not need a complex address translation mechanism as that provided in the core. This substantiates our proposal to allow an accelerator to manage its own address translation.

\section{Concluding Remarks}

We introduce \frameworkname, a comprehensive simulation framework built on the gem5 platform, specifically designed to evaluate fine-grained accelerators in high-performance systems. By integrating detailed hardware and software stacks, including a Linux software environment, Choreographer enables accurate modeling of system-level interactions and overheads. Case studies on a data-aware prefetcher and a quicksort accelerator demonstrated the framework's versatility, achieving up to 1.88x speedup and over 2x speedup, respectively. These results highlight Choreographer's ability to model and optimize fine-grained accelerator designs effectively.

\frameworkname \space leverages gem5’s strengths in full-system simulation while extending its capabilities to address the unique challenges of fine-grained task offloading. Its modularity, extensibility, and realistic system-level simulations make it a valuable tool for researchers and practitioners aiming to design efficient, latency-sensitive accelerators. Future work will explore multi-threaded scenarios and dynamic optimization techniques to further enhance the applicability of the framework.

\bibliographystyle{IEEEtranS}
\bibliography{references.bib}

\end{document}